\documentclass[aps,twocolumn,superscriptaddress,showpacs]{revtex4}

\usepackage[dvips]{graphicx} 
\usepackage{color}           
\usepackage{amsmath,amssymb,amsfonts}
\usepackage{bm}
\newcommand{\nc}{\newcommand}           
\nc{\vc}[1]     {\mbox{\boldmath $#1$}} 
\nc{\mapleft}[1]{                       
 \smash{\mathop{                      %
  \hbox to 0.90cm{\rightarrowfill} }\limits_{#1}}}

\nc{\mydraft}	{\setlength{\topmargin}{-1.5cm}}
\mydraft

\begin{document}

\title{Five-body resonances in $^8$He and $^8$C using the complex scaling method}

\author{Takayuki Myo\footnote{takayuki.myo@oit.ac.jp}}
\affiliation{General Education, Faculty of Engineering, Osaka Institute of Technology, Osaka, Osaka 535-8585, Japan}
\affiliation{Research Center for Nuclear Physics (RCNP), Osaka University, Ibaraki 567-0047, Japan}

\author{Myagmarjav Odsuren\footnote{odsuren@seas.num.edu.mn}}
\affiliation{School of Engineering and Applied Sciences, Nuclear Research Centre, National University of Mongolia, Ulaanbaatar 210646, Mongolia}

\author{Kiyoshi Kat\=o\footnote{kato@nucl.sci.hokudai.ac.jp}}
\affiliation{Nuclear Reaction Data Centre, Faculty of Science, Hokkaido University, Sapporo 060-0810, Japan}

\date{\today}

\begin{abstract}
  We study many-body resonances in the neuron-rich $^8$He and the mirror proton-rich $^8$C using the $^4$He+$N$+$N$+$N$+$N$ five-body model
  with the isospin $T=2$ system.
Resonances are described with the complex energy eigenvalues as the Gamow states using the complex scaling method.
In $^8$He, we obtain five states, in which four states are resonances, and in $^8$C all five states are resonances.
We discuss the isospin-symmetry breaking dynamically induced by the Coulomb interaction
in the energy spectra and decay widths of the resonances in two mirror nuclei.
We predict the resonance energies and decay widths for the future experiments.
We also investigate the configurations of valence nucleons above $^4$He in two nuclei with the $jj$ coupling scheme and all the states dominantly have the $p$-shell configurations.
From the configuration mixing, $^8$He and $^8$C give the similar results, which indicates the good symmetry in two nuclei.
\end{abstract}



\maketitle 

\section{Introduction}

Experiments using radioactive beams have brought the development of physics of unstable nuclei.
Neutron halo structure is one of new phenomena of nuclear structures appearing in the drip-line nuclei, such as $^6$He, $^{11}$Li, and $^{11}$Be \cite{tanihata85,tanihata13}.
One of the characteristic features in unstable nuclei is a weak binding nature of a last few nucleons and this property causes many states observed above the particle thresholds. 
This means that spectroscopy of resonances of unstable nuclei provides the important information to understand the nuclear structure.
There are two sides of neutron-rich and proton-rich in unstable nuclei with a large isospin and the comparison of the structures of these mirror systems is also interesting
to understand the isospin-symmetry property with a large isospin system.

So far, many experiments have been performed for neutron-rich $^8$He \cite{korsheninnikov93,iwata00,meister02,chulkov05,skaza07,mueller07,golovkov09} and proton-rich $^8$C \cite{charity10,charity11},
which are in the mirror relation with the isospin $T=2$ system. 
The ground state of $^8$He has a neutron skin structure of four neutrons around $^4$He with a small separation energy of about 3 MeV.
The excited states in $^8$He are not settled yet and are considered to exist above the $^4$He+$4n$ threshold energy.
This means that the observed resonances of $^8$He can decay into the various channels of $^7$He+$n$, $^6$He+2$n$, $^5$He+3$n$, and $^4$He+4$n$.
This property of the multiparticle decays causes the difficulty to determine the energy positions of resonances in $^8$He experimentally.
Theoretically the di-neutron cluster correlation is suggested in the excited state of $^8$He \cite{enyo07}.

The ground state of the unbound nucleus $^8$C is experimentally located at 3.4 MeV above the $^4$He+$4p$ threshold energy \cite{charity11}, and the excited states of $^8$C have not yet been confirmed.
Similar to $^8$He, the $^8$C states can decay into the channels of $^7$B+$p$, $^6$Be+2$p$, $^5$Li+3$p$, and $^4$He+4$p$.
The comparison of $^8$He and $^8$C is interesting to understand effects of the Coulomb interaction in proton-rich nuclei and the nuclear isospin symmetry.

In the picture consisting of $^4$He and four valence nucleons, we analyze the He isotopes and their mirror nuclei with the $^4$He+$N+N+N+N$ five-body model \cite{myo10,myo12,myo14b}.
We solve the motion of multivalence nucleons around $^4$He in the cluster orbital shell model (COSM) \cite{suzuki88,masui06,myo07,myo11}.
The advantage of the COSM is that we can reproduce the threshold energies of the subsystems in the $A$=$8$ systems.
This aspect is important to describe the open channels for nucleon emissions and then we can treat the many-body decaying phenomena.
We describe many-body resonances applying the complex scaling method (CSM) \cite{ho83,moiseyev98,aoyama06,myo14a,myo20} 
giving the correct boundary conditions for decay channels. 
In the CSM, the wave function of resonance is obtained by solving the eigenvalue problem of the complex-scaled Hamiltonian using the $L^2$ basis functions.
Results of nuclear resonances using the CSM have been successfully shown not only for energies and decay widths, but also for spectroscopic factors and the transition strength functions by using the Green's function \cite{myo14a,myo20,myo98,myo01,suzuki05}.

In our previous works of neutron-rich He isotopes and their mirror proton-rich nuclei \cite{myo10,myo12,myo14b,myo11},
we discussed the isospin symmetry in $^7$He and $^7$B with the $^4$He+$N+N+N$ model \cite{myo11}. 
The isospin-symmetry breaking occurs in their ground states for the mixing of $2^+$ states of the $A=6$ subsystems.
This is because the relative energy distances between the $A$=$7$ states and the ``$A$=$6$''+$N$ thresholds can be different 
in $^7$He and $^7$B due to the Coulomb interaction in $^7$B.
For $A$=$8$ systems of $^8$He and $^8$C, we calculated only $0^+$ states due to the limited numerical resources to treat the large Hamiltonian matrix \cite{myo10,myo12,myo14b}.
We compared the spatial structures in the radii of two nuclei, and it is found that the Coulomb barrier prevents the valence nucleons from the spatial extension, which results in the smaller radius of $^8$C than that of $^8$He in their corresponding excited $0^+_2$ resonances.  
The same relation can also be seen between $^6$He and $^6$Be \cite{myo14b}.

In this paper, we proceed our study of many-body resonances of $^8$He and $^8$C with the $^4$He+$N+N+N+N$ five-body model.
This paper is the extension of the previous ones, in which only the $0^+$ states were investigated \cite{myo10,myo12}.
We fully calculate the other possible spin states in addition to the $0^+$ for the complete understanding of the resonance spectroscopy of two nuclei.
We predict resonances of two nuclei and examine the dominant configurations of four nucleons in each state.
These information is useful for the future experiments for two nuclei.
We also compare the configuration structures of $^8$He and $^8$C in the viewpoint of the isospin symmetry.

In Sec.~\ref{sec:method}, we explain the COSM and the CSM. 
In Sec.~\ref{sec:result}, we show the results of five-body bound and resonant states obtained in $^8$He and $^8$C.
A summary is given in Sec.~\ref{sec:summary}.

\section{Method}\label{sec:method}

\subsection{Cluster orbital shell model}

\begin{figure}[t]
\centering
\includegraphics[width=4.5cm,clip]{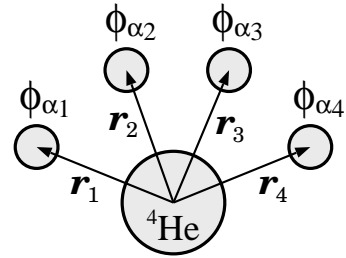}
\caption{Spatial coordinates of the $^4$He+$N$+$N$+$N$+$N$ system in the COSM.}
\label{fig:COSM}
\end{figure}

We explain the COSM with the $^4$He+$N$+$N$+$N$+$N$ five-body systems for $^8$He and $^8$C.
Motion of four nucleons around $^4$He is solved in the COSM.
The relative coordinates of four nucleons are $\{\vc{r}_i\}$ with $i=1,\ldots,4$ as are shown in Fig.~\ref{fig:COSM}.
We employ the common Hamiltonian used in the previous studies~\cite{myo10,myo12,myo14b};
\begin{eqnarray}
	H
&=&	t_0+ \sum_{i=1}^4 t_i - T_G + \sum_{i=1}^4 v^{\alpha N}_i + \sum_{i<j}^4 v^{NN}_{ij}
    \\
&=&	\sum_{i=1}^4 \left( \frac{\vc{p}^2_i}{2\mu} + v^{\alpha N}_i \right) + \sum_{i<j}^4 \left( \frac{\vc{p}_i\cdot \vc{p}_j}{4m} + v^{NN}_{ij} \right) .
    \label{eq:Ham}
\end{eqnarray}
The kinetic energy operators $t_0$, $t_i$, and $T_G$ are for $^4$He, a valence nucleon and the center-of-mass parts, respectively.
The operator $\vc{p}_i$ is the relative momentum between $^4$He and a valence nucleon
with the reduced mass $\mu$=$4m/5$ using a nucleon mass $m$.
The $^4$He--nucleon interaction $v^{\alpha N}$ has the nuclear and Coulomb parts.
The nuclear part is the microscopic Kanata-Kaneko-Nagata-Nomoto potential \cite{aoyama06,kanada79},
which reproduces the $^4$He-nucleon scattering data.
The Coulomb part is obtained by folding the density of $^4$He with the $(0s)^4$ configurations.
For nucleon-nucleon interaction $v^{NN}$, we use the Minnesota nuclear potential \cite{tang78}
and the point Coulomb interaction between protons.

We explain the COSM wave function.
We assume the $^4$He wave function $\Phi(^4{\rm He})$ with the $(0s)^4$ configuration in a harmonic oscillator basis state.
The range parameter of the $0s$ state is 1.4 fm reproducing the charge radius of $^4$He.
We expand the wave functions of the $^4$He+$N$+$N$+$N$+$N$ system using the COSM configurations \cite{masui06,myo10,myo12}.
The total wave function of a nucleus $\Psi^J$ with total spin $J$ is given in the form of the linear combination of the COSM configuration $\Psi^J_c$ as
\begin{eqnarray}
    \Psi^J
&=& \sum_c C^J_c \Psi^J_c,
    \label{WF0}
    \\
    \Psi^J_c
 &=& {\cal A}' \left\{ \Phi(^4{\rm He}), \Phi^J_c \right\},
    \\
    \Phi^J_c
&=& {\cal A} \left[ \left[ \phi_{\alpha_1}, \phi_{\alpha_2} \right]_{j_{12}},\left[ \phi_{\alpha_3},\phi_{\alpha_4} \right]_{j_{34}} \right]_J .
    \label{WF1}
\end{eqnarray}
The single-particle wave function is $\phi_{\alpha}(\vc{r})$ with the quantum number $\alpha$ as the set of $\{n,\ell,j\}$ in the $jj$ coupling scheme.
The index $n$ is to distinguish the different radial component and $\ell$ is the orbital angular momentum with $j=[\ell,1/2]$.
The spins of $j_{12}$ and $j_{34}$ are for the coupling of two nucleons.
The operators ${\cal A'}$ and ${\cal A}$ are for the antisymmetrization between $^4$He and a valence nucleon
and between valence nucleons, respectively.
The former effect is considered by using the orthogonality condition model \cite{aoyama06},
in which the relative $0s$ component is removed from the $\phi_{\alpha}$.
The index $c$ in Eq.~(\ref{WF0}) indicates the set of $\alpha_i$, $j_{12}$, and $j_{34}$ as $c=\{\alpha_1,\ldots,\alpha_4,j_{12},j_{34}\}$.
We take a summation of the available COSM configurations for a total spin $J$ and superpose them with the amplitude of $C^J_c$ in Eq.~(\ref{WF0}).

We calculate the Hamiltonian matrix in the COSM and solve the following eigenvalue problem
\begin{eqnarray}
  \sum_{c'}\langle \Psi^J_c |H|  \Psi^J_{c'} \rangle\, C^J_{c'} &=& E^J C_{c}^J  .
\end{eqnarray}
We obtain all the amplitudes $\{C_c^J\}$ in Eq.~(\ref{WF0}), which determine the total wave function, with the energy eigenvalue $E^J$
measured from the threshold energy of $^4$He+$N$+$N$+$N$+$N$.

The single-particle wave function $\phi_\alpha(\vc{r})$ is a function of the relative coordinate $\vc{r}$
from the center-of-mass of $^4$He to a valence nucleon as shown in Fig.~\ref{fig:COSM}.
We prepare a sufficient number of a single-particle basis function with various spatial distributions.
We expand $\phi_\alpha(\vc{r})$ by using the Gaussian functions for each single-particle orbit
\begin{eqnarray}
    \phi_\alpha(\vc{r})
&=& \sum_{k=1}^{N_{\ell j}} d^k_{\alpha}\ u_{\ell j}(\vc{r},b_{\ell j}^k)\, ,
    \label{spo}
    \\
    u_{\ell j}(\vc{r},b_{\ell j}^k)
&=& N_k \, r^{\ell} e^{-(r/b_{\ell j}^k)^2/2}\, [Y_{\ell}(\hat{\vc{r}}),\chi^\sigma_{1/2}]_{j}\, ,
    \label{Gauss}
	\\
    \langle \phi_\alpha | \phi_{\alpha'} \rangle 
&=& \delta_{\alpha,\alpha'}
~=~ \delta_{n,n'}\, \delta_{\ell,\ell'}\, \delta_{j,j'}.
    \label{Gauss2}
\end{eqnarray}
The index $k$ is to specify the Gaussian functions with the range parameter $b_{\ell j}^k$
with $k=1,\ldots, N_{\ell j}$ for radial correlation. The normalization factor is given by $N_k$. 
The coefficients $\{d^k_{\alpha}\}$ in Eq.~(\ref{spo}) are obtained from the orthogonal property of the basis states $\phi_\alpha$ in Eq.~(\ref{Gauss2}).
The length parameters $b_{\ell j}^k$ are chosen in geometric progression.
Basis number $N_{\ell j}$ for $\phi_\alpha$ is determined to converge the numerical results and we use 14 Gaussian functions at most in the ranges of $b_{\ell j}^k$ from 0.2 fm to around 50 fm.
We expand each of the COSM configuration $\Phi^J_c$ in Eq.~(\ref{WF1}) using a finite number of single-particle basis states $\phi_\alpha$ for each nucleon.
After solving the eigenvalue problem of Hamiltonian, we obtain the energy eigenvalues, which are discretized for bound, resonant and continuum states.

For the single-particle orbits $\phi_\alpha$, we consider the basis states with the orbital angular momenta $\ell\le 2$, and this condition gives the two-neutron energy of $^6$He($0^+$) in the accuracy of 0.3 MeV from the convergent calculation with a large $\ell$.
In this paper, we adopt the 173.7 MeV of the repulsive strength of the Minnesota potential from the original 200 MeV to fit the two-neutron separation energy of $^6$He with the experimental one of 0.975 MeV.
This treatment nicely works to reproduce the energy levels of He isotopes and their mirror nuclei systematically \cite{myo14a}.

\subsection{Complex scaling method}

We explain the CSM to treat resonances and continuum states in the many-body system \cite{ho83,moiseyev98,aoyama06,myo14a,myo20}.
The resonances are defined to be the eigenstates having the complex eigenenergies as the Gamow states with the outgoing boundary condition, and the continuum states are orthogonal to the resonances.
In the CSM, all the relative coordinates $\{\vc{r}_i\}$ in the $^4$He+$N$+$N$+$N$+$N$ system as shown in Fig. \ref{fig:COSM}, are transformed using a common scaling angle $\theta$ as
\begin{eqnarray}
  \vc{r}_i \to \vc{r}_i\, e^{i\theta}.
\end{eqnarray}
The Hamiltonian in Eq.~(\ref{eq:Ham}) is transformed into the complex-scaled Hamiltonian $H_\theta$, and the complex-scaled Schr\"odinger equation is written as
\begin{eqnarray}
	H_\theta\Psi^J_\theta
&=&     E^J \Psi^J_\theta .
	\label{eq:eigen}
        \\
    \Psi^J_\theta
&=& \sum_c C^J_{c,\theta} \Psi^J_c.
    \label{WF_CSM}
\end{eqnarray}
The eigenstates $\Psi^J_\theta$ are determined by solving the eigenvalue problem in Eq.~(\ref{eq:eigen}).
In the total wave function, the $\theta$ dependence is included in the expansion coefficients $C_{c,\theta}^J$ in Eq.~(\ref{WF_CSM}), which can be complex numbers in general.
We obtain the energy eigenvalues $E^J$ of bound and unbound states on a complex energy plane, which are governed by the ABC theorem \cite{ABC}.
From the ABC theorem, the asymptotic boundary condition of resonances is transformed to the damping behavior. 
This proof is mathematically general in many-body systems.
The boundary condition of the resonances in the CSM makes it possible to use the numerical method to obtain the bound states in the calculation of resonances.
In the CSM, the Riemann branch cuts are commonly rotated down by $2\theta$ in the complex energy plane, 
in which each of the branch cuts starts from the corresponding threshold energy.
On the other hand, the energy eigenvalues of bound and resonant states are independent of $\theta$ from the ABC theorem.
We identify the resonances with complex energy eigenvalues as $E=E_r-i\Gamma/2$, 
where $E_r$ and $\Gamma$ are the resonance energies and the decay widths, respectively. 
The scaling angle $\theta$ is determined in each resonance to give the stationary point of the energy eigenvalue on the complex energy plane.

In the CSM, resonance wave functions can be expanded in terms of the $L^2$ basis functions because of the damping boundary condition,
and the amplitudes of resonances are normalized with the condition of $\sum_{c} \big(C^J_{c,\theta}\big)^2=1$.
It is noted that the Hermitian product is not adopted due to the bi-orthogonal property of the adjoint states \cite{ho83,moiseyev98,berggren68}.
Hence the components of the COSM configurations $\big(C^J_{c,\theta}\big)^2$ can be a complex number
and are independent of the scaling angle $\theta$ when we obtain the converging solutions of resonances \cite{myo20}.

\section{Results}\label{sec:result}

\subsection{Energy spectra of He isotopes and mirror nuclei}

We discuss the resonances in $^8$He and $^8$C in the COSM.
The energy eigenvalues obtained in two nuclei are listed in Tables \ref{ene_8He} and \ref{ene_8C}, which are measured from the thresholds of $^4$He+$N+N+N+N$.
We obtain five states in each nucleus, and only the ground state of $^8$He is a bound state and the others are resonances.

For the ground state of $^8$He, the relative energy from $^4$He is 3.22 MeV and close to the experimental value of 3.11 MeV.
The matter and charge radii of $^8$He are 2.52 and 1.92 fm, respectively, which also reproduce the experiments \cite{mueller07}.
The detailed analysis of this state was reported in the previous analysis \cite{myo10,myo14b}.
For the $2^+_1$ resonance of $^8$He, we obtain the relative energy from $^4$He is 0.32 MeV and the decay width $\Gamma$ of 0.66 MeV, which are consistent to the old experimental value of
the corresponding energy of 0.46$\pm$0.12 MeV and $\Gamma$=0.5 $\pm$ 0.35 MeV \cite{korsheninnikov93}.
In the two Tables \ref{ene_8He} and \ref{ene_8C}, it is found that the energies of the $^8$C states are entirely located higher than those of $^8$He and the decay widths becomes larger for resonances
in $^8$C than $^8$He. This indicates the dynamical isospin-symmetry breaking induced by the Coulomb interaction for valence protons in $^8$C.
The detailed configurations of each state will be discussed later.

\begin{table}[t]
  \caption{Energy eigenvalues of $^8$He measured from the $^4$He+$n$+$n$+$n$+$n$ threshold energy in units of MeV.
    Data of $0^+_{1,2}$ are taken from Ref. \cite{myo10}. The values in the square brackets are the experimental ones of $0^+_1$
    and $2^+_1$ \cite{korsheninnikov93}. }
\label{ene_8He}
\centering
\begin{ruledtabular}
\begin{tabular}{c|ccc}
          & Energy~(MeV)      &  Decay width~(MeV)       \\ \hline
 $0^+_1$  & $-$3.22 [$-$3.11] &  ---               \\
 $0^+_2$  & 3.07              &  3.19              \\
 $1^+  $  & 1.65              &  3.57              \\
 $2^+_1$  &~~0.32~[0.46$\pm$0.12]        & ~~0.66~[0.5$\pm$0.35]   \\
 $2^+_2$  & 4.52              &  4.39              \\
\end{tabular}
\end{ruledtabular}
\end{table}
\begin{table}[t]
\caption{Energy eigenvalues of $^8$C measured from the $^4$He+$p$+$p$+$p$+$p$ threshold energy in units of MeV.
  Data of $0^+_{1,2}$ are taken from Ref. \cite{myo12}.
  The values in the square brackets are the experimental ones for the ground state \cite{charity11}. }
\label{ene_8C}
\centering
\begin{ruledtabular}
\begin{tabular}{c|ccc}
          & Energy~(MeV)      &  Decay width~(MeV)   \\ \hline
 $0^+_1$  & 3.32~[3.449(30)]  &  0.072~[0.130(50)] \\
 $0^+_2$  & 8.88              &  6.64              \\
 $1^+  $  & 7.89              &  7.28              \\
 $2^+_1$  & 6.38              &  4.29              \\
 $2^+_2$  & 9.70              &  9.10              \\
\end{tabular}
\end{ruledtabular}
\end{table}

\begin{figure}[t]
\centering
\includegraphics[width=8.5cm,clip]{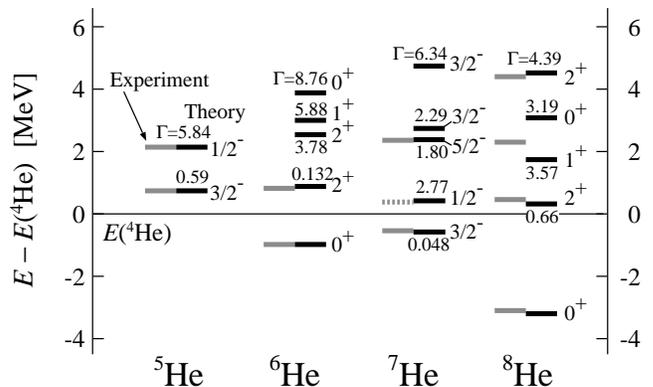}
\caption{Energy levels of $^{4-8}$He measured from the $^4$He energy. Units are in MeV.
Black and gray lines are the values of theory and experiments, respectively. Small numbers indicate the decay widths $\Gamma$ of resonances.
For $^7$He($1/2^-$), the experimental data are taken from Ref. \cite{skaza06}.
For $^8$He, the experimental data are taken from Ref. \cite{golovkov09}.}
\label{fig:He}
\end{figure}

\begin{figure}[th]
\centering
\includegraphics[width=8.5cm,clip]{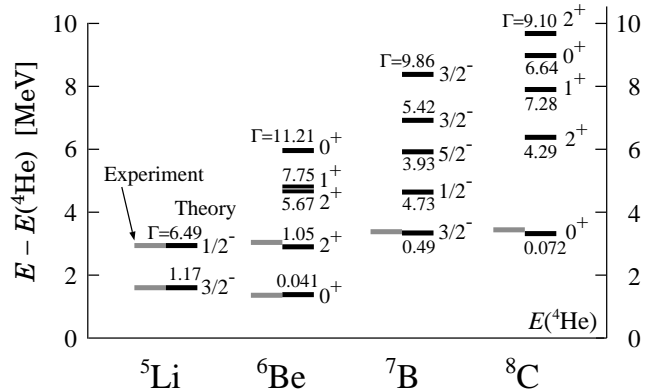}
\caption{Energy levels of $^5$Li, $^6$Be, $^7$B, and $^8$C. Units are in MeV.
Notations are the same as shown in Fig.~\ref{fig:He}.}
\label{fig:mirror}
\end{figure}

\begin{figure}[th]
\centering
\includegraphics[width=8.5cm,clip]{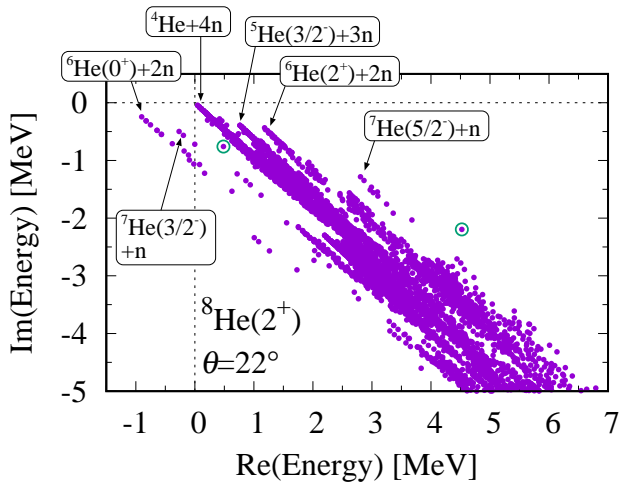}
\caption{
  Energy eigenvalue distribution of $^8$He ($2^+$) measured from the $^4$He+$n$+$n$+$n$+$n$ threshold energy in the complex energy plane,
  where scaling angle is 22$^\circ$. Units are in MeV.
  The energies with double circles indicate the $2^+_1$ and $2^+_2$ resonances.
  Several groups of the continuum states are shown with their configurations.}
\label{fig:CSM}
\end{figure}

\begin{figure}[th]
\centering
\includegraphics[width=5.0cm,clip]{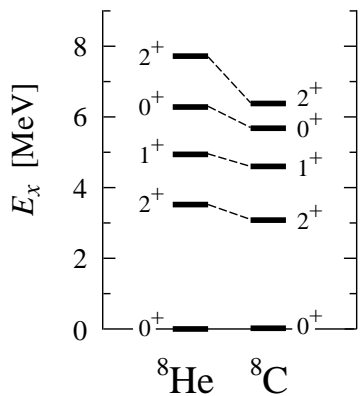}
\caption{Comparison of the excitation energy spectra of $^8$He and $^8$C in the units of MeV.}
\label{fig:excite}
\end{figure}

We show the systematic behavior of energy levels of $^{4-8}$He in Fig. \ref{fig:He} 
and their mirror nuclei of $^5$Li, $^6$Be, $^7$B, and $^8$C in Fig. \ref{fig:mirror}.
In these levels, new results in the present analysis are the $1^+$ and $2^+$ states of $^8$He and $^8$C.

In Fig. \ref{fig:CSM}, we show the example of the calculated energy eigenvalues of $^8$He($2^+$) in the CSM using $\theta=22^\circ$,
which gives the stationary condition for the energy of the $2^+_1$ state.
It is confirmed that in addition to the $2^+_{1,2}$ resonances which are clearly confirmed,
many kinds of the threshold energy positions and the corresponding continuum states are obtained with discretized spectra.
In particular, it is found that the $2^+_1$ resonance is located near the threshold energies of various continuum states as shown in Fig. \ref{fig:CSM}.
This indicates that one should carefully distinguish the components of resonance and continuum states in the observables.
It is interesting to investigate the effects of resonances and continuum states on the cross section in the future,
which can be performed by using the Green's function with complex scaling \cite{myo98,myo01,myo14a}. 

It is meaningful to discuss the isospin symmetry between $^8$He and $^8$C for four valence neutrons and protons above $^4$He with $T=2$.
We compare the excitation energy spectra of two nuclei measured from the ground states using their resonance energies in Fig.~\ref{fig:excite}.
The level orders are the same in two nuclei, but, the level spacing is smaller in $^8$C than that of $^8$He,
indicating a dynamical isospin-symmetry breaking induced by the Coulomb interaction for protons.
This result is also related to the fact that the resonances in $^8$C have larger decay widths than those of $^8$He as shown in Tables \ref{ene_8He} and \ref{ene_8C}.
When we compare the direct distance between the complex energy eigenvalues of $E_r-i\Gamma/2$, for example, $0^+_1$ and $2^+_2$, 
$^8$He and $^8$C give 8.05 and 7.84 MeV, respectively, and these values are close to each other.
In this sense, we confirm the similar energy spectra in two nuclei for complex energy eigenvalues in the complex energy plane.

\subsection{Configurations of $^8$He and $^8$C}

\begin{table}[tb]
  \caption{Dominant parts of the squared amplitudes $(C_{c,\theta}^J)^2$ of the $0^+_1$ states of $^8$He and $^8$C.
The values are taken from Ref.\cite{myo14b}.
  }
\label{comp8_1}
\centering
\begin{ruledtabular}
\begin{tabular}{c|ccc}
Configuration              &  $^8$He($0^+_1$) & $^8$C($0^+_1$)  \\ \hline
 $(p_{3/2})^4$             &  0.860           & $0.878-i0.005$  \\
 $(p_{3/2})^2(p_{1/2})^2$  &  0.069           & $0.057+i0.001$  \\
 $(p_{3/2})^2(1s_{1/2})^2$ &  0.006           & $0.010+i0.003$  \\
 $(p_{3/2})^2(d_{3/2})^2$  &  0.008           & $0.007+i0.000$  \\
 $(p_{3/2})^2(d_{5/2})^2$  &  0.042           & $0.037+i0.000$  \\
\end{tabular}
\end{ruledtabular}
\end{table}

\begin{table}[tb]
  \caption{Dominant parts of the squared amplitudes $(C_{c,\theta}^J)^2$ of the $0^+_2$ states of $^8$He and $^8$C.
The values are taken from Ref.\cite{myo14b}.
}
\label{comp8_2}
\centering
\begin{ruledtabular}
\begin{tabular}{c|ccc}
Configuration              &  $^8$He($0^+_2$) & $^8$C($0^+_2$)  \\ \hline
 $(p_{3/2})^4$             &  $0.020-i0.009$  & $0.044+i0.007$  \\
 $(p_{3/2})^2(p_{1/2})^2$  &  $0.969-i0.011$  & $0.934-i0.012$  \\
 $(p_{3/2})^2(1s_{1/2})^2$ & $-0.010-i0.001$  & $-0.001+i0.000$ \\
 $(p_{3/2})^2(d_{3/2})^2$  &  $0.018+i0.022$  & $0.020+i0.003$  \\
 $(p_{3/2})^2(d_{5/2})^2$  &  $0.002+i0.000$  & $0.002+i0.001$  \\
\end{tabular}
\end{ruledtabular}
\end{table}

\begin{table}[th]
\caption{Dominant parts of the squared amplitudes $(C_{c,\theta}^J)^2$ of the $1^+$ states of $^8$He and $^8$C.}
\label{comp8_3}
\centering
\begin{ruledtabular}
\begin{tabular}{c|ccc}
Configuration                            &  $^8$He($1^+$)   & $^8$C($1^+$)    \\ \hline
 $(p_{3/2})^3(p_{1/2})$                  &  $0.949-i0.027$  & $0.962+i0.008$  \\
 $(p_{3/2})(p_{1/2})^2(\tilde{p}_{1/2})$ &  $0.017+i0.020$  & $0.011-i0.011$  \\
 $(p_{3/2})(p_{1/2})(d_{5/2})^2$         &  $0.013+i0.006$  & $0.009+i0.002$  \\
\end{tabular}
\end{ruledtabular}
\end{table}

\begin{table}[th]
\caption{Dominant parts of the squared amplitudes $(C_{c,\theta}^J)^2$ of the $2^+_1$ states of $^8$He and $^8$C.}
\label{comp8_4}
\centering
\begin{ruledtabular}
\begin{tabular}{c|ccc}
Configuration                             &  $^8$He($2^+_1$) & $^8$C($2^+_1$)  \\ \hline
 $(p_{3/2})^3(p_{1/2})$                   &  $0.922-i0.000$  & $ 0.922+i0.017$  \\
 $(p_{3/2})^2(p_{1/2})^2$                 &  $0.021-i0.009$  & $ 0.035-i0.028$  \\
 $(p_{3/2})(p_{1/2})^2(\tilde{p}_{1/2})$  &  $0.015+i0.009$  & $-0.009+i0.014$  \\
 $(p_{3/2})(p_{1/2})(d_{5/2})^2$          &  $0.010+i0.003$  & $ 0.008+i0.003$  \\
\end{tabular}
\end{ruledtabular}
\end{table}

\begin{table}[th]
\caption{Dominant parts of the squared amplitudes $(C_{c,\theta}^J)^2$ of the $2^+_2$ states of $^8$He and $^8$C.}
\label{comp8_5}
\centering
\begin{ruledtabular}
\begin{tabular}{c|ccc}
Configuration                             &  $^8$He($2^+_2$) & $^8$C($2^+_2$)  \\ \hline
$(p_{3/2})^2(p_{1/2})^2$                  &  $0.908+i0.015$  & $ 0.955+i0.052$  \\
$(p_{3/2})^3(p_{1/2})$                    &  $0.026-i0.011$  & $ 0.035-i0.031$  \\
 $(p_{3/2})(p_{1/2})^2(\tilde{p}_{1/2})$  &  $0.032-i0.006$  & $-0.017+i0.006$  \\
$(p_{3/2})^2(d_{3/2})^2$                  &  $0.032+i0.004$  & $ 0.010-i0.024$  \\
\end{tabular}
\end{ruledtabular}
\end{table}

We discuss the configurations of four valence nucleons in the five states of $^8$He and $^8$C.
For $0^+_{1,2}$, we already discussed the results in Refs. \cite{myo10,myo12,myo14b}, hence
we only show the values in Tables \ref{comp8_1} and  \ref{comp8_2} for reference.
New results are $1^+$ and $2^+_{1,2}$ as shown in Tables \ref{comp8_3}, \ref{comp8_4}, and \ref{comp8_5}, respectively.
We show the dominant parts of the squared amplitude $(C_{c,\theta}^J)^2$ of the COSM configurations in Eq.~(\ref{WF_CSM}) for each state.
In $(C_{c,\theta}^J)^2$, the magnitude of the imaginary parts are very small for all states.
In this case, we can use the real parts of $(C_{c,\theta}^J)^2$ to interpret the states in the physical meaning. 

For the $1^+$ states in two nuclei, they are dominated by the single configuration of $(p_{3/2})^3(p_{1/2})$ for four valence nucleons.
It is noted that in the configuration of $(p_{3/2})(p_{1/2})^2(\tilde{p}_{1/2})$, the $\tilde{p}_{1/2}$ state is orthogonal to the $p_{1/2}$ state.

For the $2^+_1$ states in two nuclei, they are dominated by the single configuration of $(p_{3/2})^3(p_{1/2})$ for four valence nucleons, which are the same as the $1^+$ results.
For the $2^+_2$ states, they are dominated by the single configuration of $(p_{3/2})^2(p_{1/2})^2$ for four valence nucleons and this single-particle configuration is the same results as the $0^+_2$ cases.
From these configuration results, when we see the five states of $^8$He and $^8$C,
the valence nucleons above $^4$He are dominantly in the $p$-shell configurations and the mixing of $sd$-shell configurations is small.

When we compare the amount of the configuration mixings among the five states of $^8$He and $^8$C, 
only their ground $0^+_1$ states show the mixing stronger than other four states in each nucleus, as shown in Table \ref{comp8_1}.
Namely, the ground states are the most correlated ones.
This is because for $^8$He, the ground state is a bound state and for $^8$C, the ground state is the resonance with a small decay width of 0.07 MeV. 
Hence, the spatial distributions of four valence nucleons in their ground states are considered to be more compact than other four resonances in each nucleus. 
This property works to enhance the couplings between different configurations in the interaction region of four valence nucleons.
This point was discussed in Ref. \cite{myo14b} by comparing the radius of valence nucleons in the $0^+_{1,2}$ states of two nuclei. 
If we adopt the bound-state approximation with $\theta=0$, namely, without the CSM,
we can obtain the $2^+_1$ state of $^8$He with the positive energy of 0.36 MeV from the $^4$He+$4n$ threshold,
and the mixings of the configurations of $(p_{3/2})^3(p_{1/2})$ and $(p_{3/2})^2(p_{1/2})^2$ are 0.89 and 0.04,
which shows the enhancement of the configuration mixing from the converging values shown in Table \ref{comp8_4} with the CSM.
This fact indicates the importance of the correct treatment of the boundary condition in the analysis of the resonances.

We remark on the experimental situation of the resonances of $^8$He and $^8$C.
For $^8$He, $2^+_1$ has been reported in some experiments with the consistent energy region \cite{korsheninnikov93,golovkov09}, but
$0^+_1$, $1^+$, and $2^+_2$ are not settled yet, although the possible signature has been reported \cite{golovkov09}.
For $^8$C, the only ground $0^+_1$ state has been reported \cite{charity11}.
It is interesting to compare the present theoretical predictions with the experimental observations
to get knowledge of the isospin-symmetry breaking in drip-line nuclei. 

\newpage 
\section{Summary}\label{sec:summary}

We investigated the resonances of $^8$He and $^8$C using the $^4$He+$N+N+N+N$ five-body model for neutron-rich and proton-rich nuclei.
We use the cluster orbital shell model to describe the multi-nucleon motion around $^4$He in the weakly bound and unbound states.
We also use the complex scaling method to treat many-body resonances under the correct boundary condition.

We found five states in both $^8$He and $^8$C, which are resonances except for the ground state of $^8$He.
We obtained the resonance energies and decay widths from the complex energy eigenvalues of the resonance poles.
The dynamical isospin-symmetry breaking is confirmed in the energy spectra of $^8$He and $^8$C including their decay widths,
induced by the Coulomb interaction for protons in $^8$C.
The obtained states are dominantly explained in the $p$-shell configurations, and 
the resonances with larger decay widths tend to be dominated by the single configuration in the $jj$-coupling picture.
The excited resonances obtained in the present paper of $^8$He and $^8$C are the prediction for the future experiments.

This paper is the extension of the previous systematic study of neutron-rich He isotopes and their proton-rich mirror nuclei.
In the future, the detailed analysis of the each resonance will be performed for $^8$He and $^8$C.
For $^8$He, there are many kinds of open channels in the low-energy region, and
the transition strength from the ground state is interesting to investigate the effect of not only resonances, but also non-resonant continuum states related to the open channels.
In the calculation of the strength functions, the Green's function with complex scaling is useful \cite{myo98,odsuren14,odsuren15,odsuren17}.
The isospin-symmetry analysis is also an interesting aspect in two nuclei near the drip lines.

\section*{Acknowledgments}
This work was supported by JSPS KAKENHI Grants No. JP18K03660 and No. JP20K03962.
Numerical calculations were partly achieved through the use of OCTOPUS at the Cybermedia Center, Osaka University.
\clearpage

\section*{References}
\def\JL#1#2#3#4{ {{\rm #1}} \textbf{#2}, #4 (#3)}  
\nc{\PR}[3]     {\JL{Phys. Rev.}{#1}{#2}{#3}}
\nc{\PRC}[3]    {\JL{Phys. Rev.~C}{#1}{#2}{#3}}
\nc{\PRA}[3]    {\JL{Phys. Rev.~A}{#1}{#2}{#3}}
\nc{\PRL}[3]    {\JL{Phys. Rev. Lett.}{#1}{#2}{#3}}
\nc{\NP}[3]     {\JL{Nucl. Phys.}{#1}{#2}{#3}}
\nc{\NPA}[3]    {\JL{Nucl. Phys.}{A#1}{#2}{#3}}
\nc{\PL}[3]     {\JL{Phys. Lett.}{#1}{#2}{#3}}
\nc{\PLB}[3]    {\JL{Phys. Lett.~B}{#1}{#2}{#3}}
\nc{\PTP}[3]    {\JL{Prog. Theor. Phys.}{#1}{#2}{#3}}
\nc{\PTPS}[3]   {\JL{Prog. Theor. Phys. Suppl.}{#1}{#2}{#3}}
\nc{\PRep}[3]   {\JL{Phys. Rep.}{#1}{#2}{#3}}
\nc{\JP}[3]     {\JL{J. of Phys.}{#1}{#2}{#3}}
\nc{\PPNP}[3]   {\JL{Prog. Part. Nucl. Phys.}{#1}{#2}{#3}}
\nc{\PTEP}[3]   {\JL{Prog. Theor. Exp. Phys.}{#1}{#2}{#3}}
\nc{\andvol}[3] {{\it ibid.}\JL{}{#1}{#2}{#3}}

\end{document}